\documentclass[twocolumn,showpacs,amsmath,epsfig,apsrev]{revtex4}
\usepackage{graphicx}
\usepackage{graphics}
\usepackage{dcolumn}
\usepackage{bm}
\usepackage{verbatim}

\begin{document}
\title{Molecular nanomagnets as quantum simulators}
\author{P. Santini$^1$, S. Carretta$^1$, F. Troiani$^2$ and G. Amoretti$^1$.}
\address{$^1$Dipartimento di Fisica, Universit\`a di
Parma, Viale G. P. Usberti 7/A, I-43124 Parma, Italy}
\address{$^2$Istituto Nanoscienze-CNR, S3, via G. Campi 213/a, I-41125 Modena, Italy}

\date{\today}

\begin{abstract}
\noindent
Quantum simulators are controllable systems that can be used to simulate other quantum systems.
Here we focus on the dynamics of a chain of molecular qubits with interposed antiferromagnetic dimers.
We theoretically show that its dynamics can be controlled by means of uniform magnetic pulses and used
to mimic the evolution of other quantum systems, including fermionic ones.
We propose two proof-of-principle experiments, based on the simulation of the Ising model
in transverse field and of
the quantum tunneling of the magnetization in a spin-1 system.
\end{abstract}

\pacs{03.67.Ac 03.67.Lx 75.50.Xx}

\maketitle
\noindent

The simulation of quantum systems by a classical computer is intrinsically inefficient,
because the required number of bits grows exponentially with the system size.
This makes many important problems in physics and chemistry intractable.
Such limitation might be overcome by quantum simulators (QSs), whose dynamics can be controlled so as to mimic the
evolution of the target system \cite{Lloyd}.\\
The implementations of quantum simulators so far proposed essentially fall into one
of two categories \cite{ScienceReview}. In the
first one, large and globally addressable systems are used to analogically simulate
specific target Hamiltonians. In the second one, QSs consist of
few, individually addressable qubits, and the time evolution of any target system
can be discretized into a sequence of logical gates.
Here we propose a hybrid approach, where the simulation of different kinds of translationally invariant models
is performed by exploiting chains of molecular nanomagnets, manipulated by means of spatially homogeneous
magnetic fields.
As in analog QSs, our hardware consists of a potentially large array of qubits,
with a geometry reflecting that of the target system,
and no local control required.
As in digital QSs, the manipulation of the QS state is here achieved by suitable
sequences of quantum gates, performed in parallel on the whole array. This permits to
simulate a large class of models for each given geometry.\\
Our proposal exploits two classes of molecular nanomagnets \cite{gatteschi} that play two distinct roles: (effective) $S=1/2$
spins are used for encoding the qubits; these are connected through antiferromagnetic
systems with nonmagnetic ground state ($S=0$), that can be controllably excited so as to
effectively switch the coupling between the qubits.
The capability of engineering complex structures consisting of weakly coupled and
monodispersed systems has been recently demonstrated in molecular magnetism \cite{naturenano}.
The wide tuneability of both the intra- and inter-molecular interactions, combined with
the possibility of coherently driving the spin dynamics \cite{Blundell,rabi}, makes these
system suitable for both spintronics \cite{sanvito} and quantum-information \cite{LossNature,lossprb,QC2,Lehmann,Trif} applications.
In addition nanomagnets can be grafted onto surfaces without altering their properties\cite{sessoli}.\\
We start by considering different kinds of one-dimensional, translationally
invariant Hamiltonians ${\mathcal H}$ that can be mapped onto a model of $1/2$ spins ${\bf s}_i$,
with nearest-neighbour (NN) interactions and Hamiltonian $H$. Our aim is to simulate the time evolution
associated with $H$ (''target evolution" $U(t)$) by means of the proposed hardware.
The first step is to approximate $U(t)$ by the Trotter-Suzuki formula ($\hbar =1$):
\begin{eqnarray}\label{eq2}
U(t)&=&e^{-i H t} \simeq \left[ e^{-i H_{{\rm odd}}^{(2)}\tau } e^{-i H_{{\rm even}}^{(2)}\tau }
e^{-i H^{(1)} \tau } \right]^n ,
\end{eqnarray}
where $ H = H_{{\rm odd}}^{(2)} + H_{{\rm even}}^{(2)} + H^{(1)} $. The contributions $H_{{\rm odd}}^{(2)}$ ($H_{{\rm even}}^{(2)}$)
include all the two-spin terms $h_{{\rm odd}}^{(2)} ( {\bf s}_{2k-1} , {\bf s}_{2k} ) $ ($h_{{\rm even}}^{(2)} ( {\bf s}_{2k} , {\bf s}_{2k+1} ) $), while $H^{(1)} $ includes all the single-spin terms.
Since $H_{{\rm odd}}^{(2)}$, $H_{{\rm even}}^{(2)}$, and $ H_1 $ generally don't commute,
Eq. (\ref{eq2}) is only exact in the limit $\tau\equiv t/n \rightarrow 0$.
Each of the three terms in parentheses in Eq. (\ref{eq2}) can be factorized into
either single- or two-spin evolution operators.
For example, under the effect of
$ \exp{(-i H_{{\rm odd}}^{(2)} \tau)}
= \otimes_{k=1}^{N/2} \exp{[-i h_{{\rm odd}}^{(2)} ( {\bf s}_{2k-1} , {\bf s}_{2k} ) \tau ]}  $
(with $N$ the number of spins), each pair evolves in the same way, and
independently of all the others.

In order to simulate the dynamics of the spin chain, we encode its
odd- and even-numbered spins into the state of two physically
distinguishable kinds of spin qubits,
$A$ and $B$, in the quantum hardware:
$ {\bf s}_{2k-1} \rightarrow {\bf S}^A_k $ and $ {\bf s}_{2k}
\rightarrow {\bf S}^B_k $ (Fig. 1(a)). Note that hereafter we use capital letters to indicate the spins and times of the quantum simulator.
In the latter, each pair $ {\bf S}^A_k {\bf S}^B_k $
(${\bf S}^B_k {\bf S}^A_{k+1}$) is physically connected through a
spin cluster $M_{AB}$ ($M_{BA}$), whose state can be manipulated so as to
effectively switch the coupling between the qubits.
Being $M_{AB}$ spectrally distinguishable from
$M_{BA}$, it will be possible to selectively switch the $AB$ or $BA$ couplings,
still by means of spatially homogeneous em pulses (see below).
The time-evolution operator $\exp{(-i H_{{\rm odd}}^{(2)} \tau)} $,
is implemented by performing sequences of identical
single- and two-qubit operations on the pairs
${\bf S}^A_k {\bf S}^B_k$.
Such pulse sequence induces on the pair of qubits the same time-evolution
induced in the time $\tau$ by $h^{(2)}_{{\rm odd}}$ on the spin pair ${\bf s}_{2k-1} {\bf s}_{2k}$.
Analogously, sequences of pulses applied to the qubit pairs
${\bf S}^B_k {\bf S}^A_{k+1}$ reproduce the time-evolution
induced on the spin pairs ${\bf s}_{2k} {\bf s}_{2k+1}$ by the terms $h^{(2)}_{{\rm even}}$.
The contribution of the single-spin Hamiltonian, $H^{(1)}$, is simulated instead
by single-qubit rotations performed simultaneously on all the qubits.
We stress that the parameters defining $H$ can be easily varied in the simulation by appropriately choosing
the phases of the pulse sequence (see discussion below Eq. (3)).
Thanks to the translational invariance of the simulated system, the number
$N_{op}$ of operations (i.e., pulses) does not depend on the
chain length $N$. In fact, $N_{op}$ is proportional to
$n l $, where $l$ is the number of terms into which
$e^{-iH\tau}$ is factorized (see Eq. \ref{eq2}).
Besides, in many models of interest ${\mathcal H}$ (e.g., symmetric
exchange between 1/2 spins), the $A$ and $B$ units need not be physically
different.\\
{\it Switching the coupling between spin qubits --- }
Turning on and off a physical exchange interaction between molecular nanomagnets on a
ns timescale is presently unfeasible.
Here we show that the effect of a switchable coupling between two nearby spin
qubits $A$ and $B$ can be obtained by manipulating the interconnecting spin
cluster $M$ between them.
As an illustrative example, we consider the case where $M$ is a dimer, consisting of
two antiferromagnetically-coupled spins (with $ S_1^M = S_2^M = 1/2 $, see Fig. 1).
\begin{figure}
	\centering
		\includegraphics[width=0.29\textwidth]{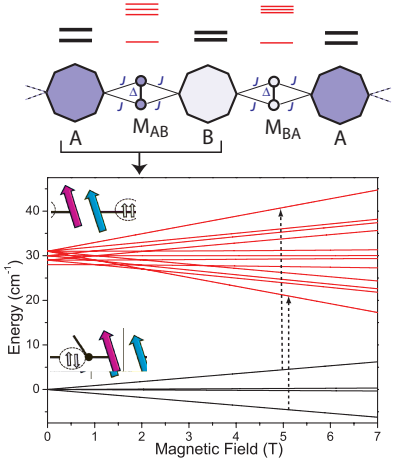}
	\caption{Top: chain of the $A$ and $B$ spin qubits, with interposed
antiferromagnetic dimers $M_{AB}$ and $M_{BA}$.
The level schemes of the noninteracting units is also illustrated.
Bottom: Magnetic field dependence of the energy levels for a $A-M-B$ system
(Eq. (\ref{eq4})), with $g_z^A=1.8$, $g_z^B=2.0$, $g_z^M=2.3$, $J=1$ cm$^{-1}$,
$\Delta=30$ cm$^{-1}$. 
When the dimer is in the singlet ground state (four lowest levels, black), the qubits behave as
if they were
non-interacting.
If the dimer is in triplet states (twelve highest levels, red), the energy levels depend on the
dimer-qubit couplings.}
	\label{fig:Fig1}
\end{figure}
\begin{figure}
	\centering
		\includegraphics[width=0.35\textwidth]{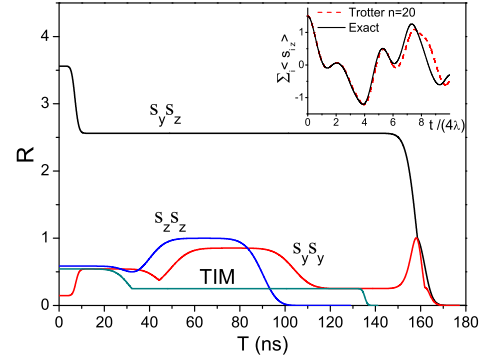}
	\caption{Simulation of the time evolution operators
$ \exp{(-i\kappa s_{1 \alpha} s_{2 \beta})} $ ($\alpha \beta = zz$, $yy$, $xy$)
for $\kappa\tau=\pi/2$. $T_f=180\,$ ns is the duration of the pulse sequence;
$R$, defined in the text, quantifies the deviation of the implemented transformation
from the target evolution operator.
The em pulses are gaussian and linearly polarized, with a peak amplitude of 50 G;
the static field is $B=5$ T. The qubits are assumed to be magnetically isotropic,
whereas for the dimer we set $g_x^A-g_x^B=g_y^A-g_y^B=1$.
$R$ is also shown for a single Trotterization step of the TIM Hamiltonian for two
spins (Eq. (4)).
Inset: time oscillations of the longitudinal average magnetization
$\langle \sum_i s_z^i\rangle $ for the TIM with $N=3$ and $\lambda = 2 b$.}
	\label{fig:Fig2}
\end{figure}
The simplest qubit-dimer-qubit unit is described by the Hamiltonian:
\begin{eqnarray}\label{eq4}
H_{AMB} &=& [H_A + H_B + H_M] + [H_{AM} + H_{BM}]  \nonumber \\ 
&=&  \big[ \mu_B B_0 \big(g_z^A S_z^A+g_z^B S_z^B+ g_z^M (S_{1 z}^M+S_{2 z}^M) \big)
  \nonumber \\
&+& \Delta {\bf S}^M_1\cdot {\bf S}^M_2 \big]+\big[J\hspace{-0.25cm}\sum_{\alpha = A,B} \sum_{i=1,2} {\bf S}_\alpha \cdot {\bf S}_i^M \big],
\end{eqnarray}
where $g_z^\chi$ are gyromagnetic factors,
and the last term is the qubit-dimer coupling, whose energy scale $J$ is typically much smaller
than $\mu_B B_0$ and $\Delta$.
A similar effective Hamiltonian results, for instance, from two Cr$_7$Ni rings linked through a Cu$_2$
dimer\cite{naturenano}, where the rings play the role of the spin qubits.
Many generalizations of Eq. (2), including less symmetric patterns of qubit-dimer exchange couplings,
do not alter the validity of the proposed scheme.
%
The field-dependence of the energies resulting from Eq. (\ref{eq4})
is depicted in Fig. 1.
The four lowest states match those of the two isolated $A$ and $B$ qubits.
In fact, as far as the dimer is in its singlet ground-state, $ \langle H_{AM} \rangle
= \langle H_{BM} \rangle = 0 $ for any two-qubit state; therefore, $A$ and $B$ are effectively
uncoupled, whereas they do communicate if the dimer is sent by an em pulse to an
excited state. Indeed, the twelve upper states have energies with fine splittings determined by $J$.\\
This level scheme can be exploited to simulate the dynamics of two generic
spins induced by a Hamiltonian $h^{(2)} = \kappa  s_{1 \alpha} s_{2 \beta} $
for any choice of $ \alpha , \beta = x,y,z $.
The operator $ \exp{(-i h^{(2)}\tau )} $ can be decomposed as follows:
\begin{eqnarray}\label{eq10}
\exp{(-i \kappa s_{1 \alpha} s_{2 \beta} \tau )} =
[u_{1 \alpha} \otimes u_{2 \beta}] e^{-i \Lambda \tau}
[u_{1 \alpha} \otimes u_{2 \beta}]^\dagger ,
\end{eqnarray}
where
$ \Lambda = \kappa s_{1 z}  s_{2 z}  $, $ u_x = (2 s_y)^{1/2} $, $ u_y = (2 s_x)^{-1/2}$, and $u_z=I$.
In the physical hardware described by Eq. (2), the
single-qubit rotations $ u_\alpha $ can be implemented by
em pulses with frequencies $B g_z^A$ and $B g_z^B$, respectively (while $M$ is left in
its ground state).
The two-qubit operator $ e^{-i \Lambda \tau} $ would in principle require
a direct interaction between A and B. Here, instead, it's implemented by inducing in
$M$ an excitation conditioned to the state of the spin qubits $A$ and $B$.
This is obtained through two simultaneous $\pi$ pulses, resonant with the gaps
indicated by arrows in Fig. 1, followed by a repetition of the two pulses
that bring the dimer back to its singlet ground state.
The value of $\kappa \tau$ of the target evolution is controlled by the phase difference
between the first and the second pair of pulses. Hence, the parameters defining the Hamiltonian we want to simulate
can be easily varied since these merely determine the phases of the pulse sequences.\\
In Fig. 2 we demonstrate the validity of this scheme by simulating the evolution
operator for some representative choices of $\alpha\beta$ in $h^{(2)}$. We start from
the 4 possible two-qubit basis states $ |\psi_i\rangle$ (corresponding to the 4 lowest
eigenstates of $H_{AMB}$, with $i=1,\dots , 4$), and we calculate their time evolution
$|\psi_i (T)\rangle$ induced by $H_{AMB}$ (Eq. (\ref{eq4})) and by the pulse
sequence. The matrix elements of the resulting transformation,
$ \tilde{U}_{ji}( T ) \equiv \langle \psi_j|\psi_i(T)\rangle$,
are compared with those of $ U(\tau )$ between the corresponding states of the system to be simulated:
the distance between the two is assessed by
$R (T) = \max_{i,j}\vert \tilde{U}_{ji} (T) - U_{ji} (\tau ) \vert^2$.
At the end of the pulse sequence ($T=T_f$),
$\tilde{U}_{ji}(T_f)$ coincides with $ U_{ji} (\tau ) $, showing that the two qubits have
actually undergone the desired unitary transformation (Fig. 2)\cite{nota}.
For the chosen, realistic parametrization, the duration of these simulations is of the
order of $10^2$ ns. This is much less than the expected decoherence times for optimally
engineered molecular qubits, of the order of several microseconds \cite{Blundell}.
At low temperatures, the coherence of each nanomagnet is limited by
the hyperfine coupling to the nuclear spins, and the noise has a
local character (i.e. each molecule interacts mainly with its own
bath of nuclear spins) \cite{szallas}.
The effects of such coupling can be partially cancelled by spin-echo
sequences. These imply the use of additional pulses, that can however
be applied in parallel to the whole array, and thus independently on
the system size.\\
{\it Quantum simulation of a spin-1/2 chain ---}
A simple proof-of-principle experiment can be performed by simulating the time evolution
of the transverse-field Ising model (TIM):
\begin{equation}
{\mathcal H}_{{\rm TIM}} \equiv H_{{\rm TIM}}  = \lambda \sum_{k=1}^{N-1} s_{k z} s_{(k+1) z} + b \sum_{k=1}^N s_{k x} ,
\end{equation}
with $s_k=1/2$.
The mapping of $H_{{\rm TIM}}$ onto the qubit chain is straightforward: $ {\bf s}_{2k-1} \rightarrow
{\bf S}_{k}^A $ and $ {\bf s}_{2k} \rightarrow {\bf S}_{k}^B $.
Performing a single Trotter step of the TIM entails simulating $e^{-i \Lambda \tau}$
with $\kappa = \lambda$, followed by a rotation of two qubits around $x$ by an angle $b\tau$.
The convergence of the simulated TIM evolution to the target evolution is shown in
Fig. 2.
For a generic value of $\lambda /b$ the TIM evolution brings the system from a factorized
initial state to multipartite entangled ones.
This is shown in the inset of Fig. 2 for the
case $N=3$. Here $ |\psi (t) \rangle $ starts from a ferromagnetic state and it evolves
passing through GHZ-like states\cite{libro}. This evolution is witnessed by oscillations of the magnetization,
whose frequencies are set by the energy gaps of the TIM.\\
{\it Simulating $S=1$ spins and the Hubbard model ---}
The simulation of Hamiltonians involving spins $s>1/2$ or fermions requires a suitable
mapping onto the qubits.
For instance, we consider a chain of spins one (${\bf t}_k$) with NN exchange interactions and single-spin
crystal-field anisotropy:
\begin{equation}\label{eq8}
{\mathcal H}_{{\rm s1}} = \lambda \sum_{k=1}^{N-1} {\bf t}_k \cdot {\bf t}_{k+1} + \sum_{k=1}^{N}
[d\; t_{k z}^2+ e(t_{k x}^2-t_{k y}^2)],
\end{equation}
which reduces to the paradigmatic Haldane model for $d=e=0$.
\begin{figure}
	\centering
		\includegraphics[width=0.4\textwidth]{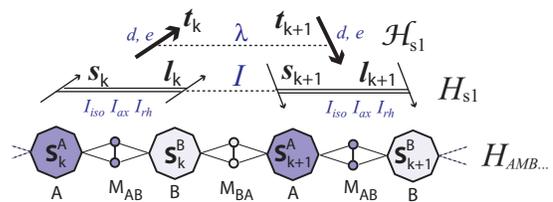}
	\caption{\label{fig4} From top to bottom: mapping of the spin-1 Hamiltonian ${\mathcal H}_{{\rm s1}}$ (Eq. (\ref{eq8})) onto a spin-1/2 one $H_{{\rm s1}}$, and encoding of $H_{{\rm s1}}$ into the spin-qubit chain $ABAB\dots$.}
\end{figure}
${\mathcal H}_{{\rm s1}}$ can be mapped onto a Hamiltonian $H_{{\rm s1}}$ of $2N$ spins $1/2$, with NN interactions.
Indeed, the dynamics of a spin-1 chain is equivalent to that of a dimerized spin-1/2 chain with twice the number of spins, provided
the isotropic exchange constant ($I_{{\rm iso}}$) (see Fig. \ref{fig4}) is ferromagnetic and dominant.
The three states $ | m_k=0,\pm 1\rangle $ of each spin $t_k = 1 $ are mapped onto
the three triplet states of the pair of spins 1/2 $ ({\bf s}_{k}, {\bf l}_{k})$, having
total spin one. By exploiting the Wigner-Eckart theorem, the crystal-field terms are mapped onto axial ($I_{{\rm ax}}$) and rhombic ($I_{{\rm rh}}$) exchange terms:
$
h_{{\rm odd}}^{(2)} ({\bf s}_{k} , {\bf l}_{k}) = I_{{\rm iso}} {\bf s}_{k} \cdot {\bf l}_{k}
+ I_{{\rm ax}} s_{k z} l_{k z} + I_{{\rm rh}} (s_{k x} l_{k x} - s_{k y} l_{k y}).
$
The exchange interaction between $ {\bf t}_k $ and $ {\bf t}_{k+1} $ in Eq. (\ref{eq8}) is
mapped instead onto a Heisenberg coupling $I$ between $ {\bf l}_{k} $ and
$ {\bf s}_{k+1} $ (Fig. \ref{fig4}):
$h_{{\rm even}}^{(2)} ({\bf l}_{k} , {\bf s}_{k+1}) = I {\bf l}_{k} \cdot {\bf s}_{k+1}$.\\
\begin{figure}
	\centering
		 \includegraphics[width=0.3\textwidth]{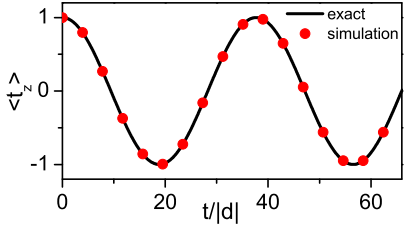}
	\caption{\label{fig5} Line: exact time evolution of $\langle t_z\rangle$ for a single $t=1$ spin with $d/e = 12$, Eq. (\ref{eq8}). The state
oscillates between $|m=1\rangle$ and $|m=-1\rangle$ due to quantum tunneling across the anisotropy barrier.
Points: time evolution simulated by a single $A-M-B$ unit initially prepared in
its ground state. The pulse sequence is set by the mapping of Eq. (\ref{eq8}) onto the Hamiltonian of two 1/2 spins (Fig. 3).
We plot the $z$-component of total spin of the $A-M-B$ unit, which could be easily extracted by measuring the magnetization of
a crystal of noninteracting units. The duration $T_f$ of the pulse sequence implementing the simulation is
about 480 ns, independently of the simulated time $t$. Note that to perform each simulation (i.e., to extract each point)
the $A-M-B$ unit has to be reinitialized to its ground state.}
\end{figure}
Having mapped ${\mathcal H}_{{\rm s1}}$ onto a chain of spins 1/2 with NN interactions, we can
now simulate its dynamics along the lines traced above. Each
spin-1/2 ${\bf s}_k$ (${\bf l}_k$) is encoded into qubit $A$ ($B$) and the operators
$\exp{(-i h_{{\rm odd}}^{(2)} \tau)}$ and $\exp{(-i h_{{\rm even}}^{(2)} \tau)}$ are mimicked as outlined above.
A simple proof-of-principle experiment would be the simulation of a single $s=1$
spin experiencing tunneling of the magnetization ($ |e| \ll |d|$). For instance, Fig. \ref{fig5}
shows the exact and simulated evolution of the magnetization in the case $d/e = 12$,
which can be monitored by measuring the total magnetization of the $A-M-B$ system.\\
The mapping of a fermionic Hamiltonian onto a spin one is generally nontrivial
\cite{Somma}.
Hereafter, we use the Jordan-Wigner representation in order to map the one-dimensional
Hubbard model, $ {\mathcal H}_{{\rm Hub}} = -t_H \sum_{k \sigma} (c^\dagger_{k \sigma} c_{k+1 \sigma} + h.c.)
+ U\sum_{k}n_{k\uparrow}n_{k\downarrow} $, onto a chain of $1/2$ spins \cite{JW}:
\begin{eqnarray}\label{eq7}
H_{{\rm Hub}} & = & NU/4 -2t_H \sum_{k=1}^{N-1} \sum_{\alpha = x,y}
( s_{k \alpha} s_{(k+1) \alpha} + l_{k \alpha} l_{(k+1) \alpha} )
 \nonumber \\
&+& U \sum_{k=1}^N s_{k z} l_{k z} + U/2\sum_{k=1}^N (s_{k z}+ l_{k z} )
\end{eqnarray}
where ${\bf s}_k$ and ${\bf l}_k$ are two families of spin-1/2 operators.
Unlike the previously considered cases, $H_{{\rm Hub}}$ is not a one-dimensional Hamiltonian
with NN interactions only, as each spin couples to three other ones. This requires to proceed in two steps:
initially we encode into the qubit pairs
$ ( {\bf S}_k^A , {\bf S}_k^B ) $
the spins
$ ( {\bf l}_k , {\bf s}_k ) $  for even $k$
and
$ ( {\bf s}_k , {\bf l}_k ) $ for odd $k$ , respectively (see Fig. \ref{fig3}).
The couplings between pairs of spins that are encoded into neighboring qubits can
be simulated as for the previously considered models. These couplings include the two $z$ terms in Eq. (\ref{eq7}) and
half of the $xy$ terms, i.e., the even transverse $s-s$ bonds and the odd transverse $l-l$ bonds.
In order to simulate the remaining transverse two-spin terms in $H_{{\rm Hub}}$, corresponding to couplings
between qubits that are initially third nearest-neighbours, we swap the state of all
the $ {\bf S}^A_k {\bf S}^B_k $ pairs. This SWAP gate can be performed by the same method used to implement two-qubit operations.
Third nearest neighbours in $H_{{\rm Hub}}$ now correspond to NN in the simulator, and the evolution induced by the associated couplings can be
simulated exactly as above.\\
\begin{figure}
	\centering
		\includegraphics[width=0.34\textwidth]{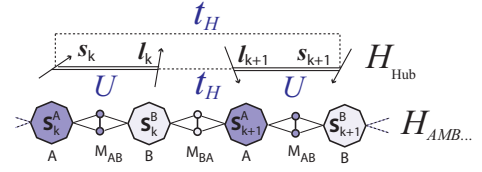}
	\caption{\label{fig3} Mapping of the Hubbard model ${\mathcal H}_{{\rm Hub}}$ onto the spin
Hamiltonian $H_{{\rm Hub}}$ involving the $s_k$ and $l_k$ 1/2 spins.}
	\label{fig:Fig3}
\end{figure}
The feasibility of our
scheme with available technology relies on the lack of local-control requirements, as
only uniform em pulses are involved.
We have illustrated the simplest possible implementation of the idea, working for
uniform or $A-B$ 1-dimensional Hamiltonians. However, extensions of this approach allow the simulation of a much larger class of Hamiltonians,
including higher-dimensional ones. One possibility is to use spin-qubit arrays that reproduce
the dimensionality $D$ of the system to simulate, with interposed nanomagnets $M$
to switch the interaction between adjacent qubits. For instance, in a square lattice $U(\tau \rightarrow 0)$ is first decomposed by the Trotter formula into two evolution operators describing a collection of identical chains (along $x$ and $y$).
The couplings within the chains can then be simulated in parallel by the method described above by using
an array of nanomagnets with rectangular symmetry. In fact, in order to selectively address the $x$ and $y$ chains, the nanomagnets $M$ switching the interaction along one direction need to be spectrally distinguishable from those operating on another direction.\\
Alternatively, it is possible to develop other simulation schemes keeping a one-dimensional topology of the hardware and the use of uniform pulses, at the cost of a more complex and less parallel algorithm. The limiting case (where all two-body terms in the Trotter-decomposed time evolution are implemented sequentially) is represented by a scheme where half of the molecular qubits are used as auxiliary units instead of logical qubits. The $D$-dimensional target Hamiltonian is mapped onto a $1D$ Hamiltonian with long-range couplings.
Since the latter prevent adopting the scheme described above for nearest-neighbor $1D$ Hamiltonians, the various two-body terms in the target Hamiltonian are simulated sequentially by the use of a control unit (in the spirit of \cite{benjamin}). Conditional excitation of the interposed dimers remains a key ingredient to induce the desired evolution, and the auxiliary units are exploited to attain local control with uniform pulses\cite{forthcoming}.\\
The capability of simulating Hamiltonians can also
be exploited in order to map experimentally accessible quantities ($O$) onto
a set of non-accessible observables ($ O' = e^{iHt} O e^{-iHt}$).
Observables corresponding to (sums of) single-qubit terms can be mapped, e.g.,
onto pair correlation functions in the case where $H$ corresponds to a dimerized
system ($ H = H_{\rm odd}^{(2)} $ or $ H = H_{\rm even}^{(2)} $), or to
higher-order correlation functions in the case of a more general $H$\cite{forthcoming}.\\
In conclusion, we have shown that arrays of molecular nanomagnets can be used
as quantum simulators of different model Hamiltonians with translational invariance
and short-range interactions.
We have proposed proof-of-principle implementations, where the means required for
manipulating the system and measuring the relevant observables can be provided by the
current technology.
Nanomagnet-dimer-nanomagnet supramolecular trimers and chains which can be exploited
to implement our scheme are also currently being synthesized \cite{private}.\\
We thank M. Affronte for useful discussions.
We acknowledge financial support from
the FP7-ICT FET Open "MolSpinQIP" project, from the PRIN of the Italian Ministery
of Research and from Fondazione Cariparma.


\begin{references}
\bibitem{nota} $\tilde{U}$ is calculated in the interaction picture for $H_{AMB} +
 H_{pulses}(t)$. The Schr\"{o}dinger picture wavefunction of the actual hardware would
 experience additional phases from the Zeeman terms in $H_{AMB}$. These can be easily
 compensated by an overall rotation of all qubits.
\end{references}
\end{document}